\documentclass{article}
\usepackage{epsfig}
\usepackage{color}
\usepackage{amsmath}
\usepackage{amssymb}

\makeatletter
\@addtoreset{equation}{section}

\makeatother

\begin{document}

\begin{flushright}
August 2012

SNUTP12-003
\end{flushright}

\begin{center}

\vspace{5cm}

{\LARGE 
\begin{center}
On Large $N$ Solution of ${\cal N}=3$ Chern-Simons-adjoint Theories
\end{center}
}

\vspace{2cm}

Takao Suyama \footnote{e-mail address : suyama@phya.snu.ac.kr}

\vspace{1cm}

{\it 
BK-21 Frontier Research Physics Division

and 

Center for Theoretical Physics, 

\vspace{2mm}

Seoul National University, 

Seoul 151-747 Korea}

\vspace{3cm}

{\bf Abstract} 

\end{center}

The planar resolvent for ${\cal N}=3$ U$(N)_k$ Chern-Simons theory coupled to an arbitrary number of adjoint matters is determined. 
Analytic continuation of the 't~Hooft coupling $t$ is analyzed. 
The eigenvalue distribution turns out to be confined in a finite region even for a large $t$. 
The vev of a Wilson loop does not exhibit an exponential growth although such a behavior would be expected for theories with classical gravity duals. 

\newpage

\vspace{1cm}

\section{Introduction}

\vspace{5mm}

Localization has been recognized as one of the most powerful techniques in analyzing supersymmetric quantum field theories which 
allows one to calculate some quantities exactly. 
The localization method of \cite{Pestun:2007rz} was applied to ${\cal N}\ge3$ Chern-Simons-matter theories on $S^3$ in \cite{Kapustin:2009kz}. 
This result, as well as its ${\cal N}=2$ generalization \cite{Jafferis:2010un}\cite{Hama:2010av}, 
can be used to calculate some physical quantities exactly. 
A recent example of such quantity can be found in \cite{Closset:2012vg}\cite{Closset:2012vp}. 

In many cases, the resulting localization 
formula for, say, the partition function is still complicated, even though it is given as a well-defined finite-dimensional 
ordinary integral. 
One way to extract some information from the formula 
is to take the planar limit of the field theory under investigation by which the integral can be evaluated exactly in terms of 
the saddle-point. 
The analysis of the saddle-point can then be performed by the traditional matrix model techniques. 
The analysis for ABJM theory \cite{Aharony:2008ug} was performed and the planar resolvent was obtained explicitly in \cite{Marino:2009jd}. 
The technique used in \cite{Marino:2009jd} was based on 
a relation of the localization formula for the partition function to a topological string theory, but this does not seem to be easily generalized 
to other Chern-Simons-matter theories. 
A simpler technique was introduced in \cite{Herzog:2010hf} (see also \cite{Suyama:2009pd} for earlier discussion) 
which could extract the leading behavior of the free energy and the 
vev of BPS Wilson loops in the large $N$ limit. 
This technique turned out to be easily generalized to various theories including ${\cal N}=2$ ones 
\cite{Martelli:2011qj}\cite{Cheon:2011vi}\cite{Jafferis:2011zi}\cite{Gulotta:2011si}\cite{Gulotta:2011aa}, 
but a systematic analysis of the $1/N$ corrections 
would not seem to be straightforward. 

Later, it was found in \cite{Fuji:2011km} that all the $1/N$ corrections in 
the partition function of ABJM theory can be explicitly summed up, resulting in a simple function. 
This surprising result was rederived in \cite{Marino:2011eh} 
by a completely different technique, that is, by using a relation of the localized partition function 
to the partition function of a particular free Fermi gas. 
This Fermi gas picture was also applied to more general theories in 
\cite{Marino:2011eh}\cite{Marino:2012az}\cite{Klemm:2012ii}. 

On the other hand, the traditional analysis of the saddle-point equations in terms of the resolvent has been developed in 
\cite{Suyama:2010hr}\cite{Santamaria:2010dm}\cite{Suyama:2011yz}, 
but the range of applicability has been rather limited. 
However, even though the progress based the Fermi gas picture is so remarkable, it seems that there are still some reasons to investigate the 
planar resolvent. 
One possible advantage would be that the planar resolvent can provide results which are exact in the 't Hooft coupling. 
This means that, if a Chern-Simons-matter theory has a string theory dual, the planar resolvent provides the complete information of the worldsheet 
theory with genus zero. 
In the Fermi gas picture, it is rather difficult to obtain some non-perturbative corrections in the sense of the worldsheet theory. 
Sooner or later, the Fermi gas picture would be improved to overcome such difficulties. 
For a while, the traditional method could be a complementary method to analyze Chern-Simons-matter theories. 

In this paper, we investigate ${\cal N}=3$ U$(N)_k$ Chern-Simons theory coupled to an arbitrary number of adjoint matters. 
We call this family of theories as ${\cal N}=3$ Chern-Simons-adjoint theories. 
An ${\cal N}=2$ version of such a theory was recently discussed in \cite{Marino:2012az} and some qualitative properties of the theories were found. 
Our theories have larger supersymmetry, and the analysis of the theories is easier. 
We determine the planar resolvent of the theories. 
The resolvent is given as an integral of a combination of elliptic theta functions. 
This integral expression of the resolvent allows us to analyze the analytic continuation of the 't~Hooft coupling and the vev 
of a BPS Wilson loop \cite{Gaiotto:2007qi}. 
The eigenvalue distribution obtained from the resolvent has a similar property observed in \cite{Marino:2012az}. 

This paper is organized as follows. 
In section \ref{MM}, we explain an ansatz for solving the saddle-point equations. 
This ansatz is applied to ${\cal N}=3$ pure Chern-Simons theory in section \ref{chapter pureCS} to check the validity of the ansatz. 
Then, ${\cal N}=3$ Chern-Simons-adjoint theories are analyzed in section \ref{chapter CSA}. 
We give integral formulas for the planar resolvent, the 't Hooft coupling and the vev of the Wilson loop as functions of the position of the eigenvalue 
distribution. 
Section \ref{discuss} is devoted to discussion. 
There are several appendices. 
Appendix \ref{G(z)} reviews a solution of a homogeneous equation studied in \cite{Eynard:1995nv}\cite{Eynard:1995zv}. 
Appendix \ref{identity} contains a proof of an identity which is used in our analysis. 
Appendix \ref{uniqueness} discusses the conditions by which the solution of the saddle-point equations is determined uniquely. 
Appendix \ref{coeff} contains some details of the calculations which determine the planar resolvent.

\vspace{1cm}

\section{Matrix models} \label{MM}

\vspace{5mm}

In this paper, we consider ${\cal N}=3$ Chern-Simons-adjoint theory, that is, ${\cal N}=3$ 
Chern-Simons theory with the gauge group U$(N)$ and the Chern-Simons level $k$ 
coupled to $2n$ matter fields in the adjoint representation of U$(N)$. 
The number of the adjoint matters is chosen to be even so as to preserve ${\cal N}=3$ supersymmetry. 
Some basic facts on ${\cal N}=3$ Chern-Simons-matter theories can be found in \cite{Gaiotto:2007qi}. 

The localization formula \cite{Kapustin:2009kz} for the partition function of ${\cal N}=3$ Chern-Simons-adjoint theory is 
\begin{equation}
Z\ =\ \int \prod_{i=1}^Ndu_i\,\exp\left[ \frac{ik}{4\pi}\sum_{i=1}^Nu_i^2 \right]
 \frac{\prod_{i<j}\sinh^2\displaystyle{\frac{u_i-u_j}2}}{\prod_{i<j}\cosh^{2n}\displaystyle{\frac{u_i-u_j}2}}. 
   \label{partition function}
\end{equation}
The normalization of $Z$ here is different from, for example, the one in \cite{Marino:2011eh}. 
This is not an important modification since the overall constant is irrelevant in the following analysis. 

The vev of a BPS Wilson loop \cite{Gaiotto:2007qi} can be also given as a finite-dimensional integral as \cite{Kapustin:2009kz} 
\begin{equation}
\langle W \rangle \ =\ \frac1Z\int \prod_{i=1}^Ndu_i\,\exp\left[ \frac{ik}{4\pi}\sum_{i=1}^Nu_i^2 \right]
 \frac{\prod_{i<j}\sinh^2\displaystyle{\frac{u_i-u_j}2}}{\prod_{i<j}\cosh^{2n}\displaystyle{\frac{u_i-u_j}2}}\cdot \frac1N\sum_{i=1}^Ne^{u_i}. 
   \label{WL}
\end{equation}

\vspace{5mm}

\subsection{Saddle-point equations} \label{SPeq}

\vspace{5mm}

In the large $N$ limit, the integral in the localization formula (\ref{partition function}) can be evaluated exactly by the saddle-point configuration $\{u_i\}$ 
which satisfies 
\begin{equation}
\frac k{2\pi}u_i\ =\ \sum_{j\ne i}\coth\frac{u_i-u_j}2-n\sum_{j}\tanh\frac{u_i-u_j}2, 
   \label{SP0}
\end{equation}
where $i$ and $j$ run from 1 to $N$, and $-ik$ is replaced with $k$, understanding that $k$ can take a complex value. 
These equations are invariant under the sign flip $u_i\to-u_i$. 
Therefore, the saddle-point configuration is expected to be symmetric under the sign flip. 
That is, an equality between two sets 
\begin{equation}
\{\,u_1, \cdots, u_N\,\}\ =\ \{\,-u_1,\cdots,-u_N\,\}
   \label{symm-u}
\end{equation}
is expected to hold. 

When $k$ is real and positive, the equations (\ref{SP0}) can be regarded as equations of the balance among forces. 
The left-hand side represents an external linear force acting on $i$-th eigenvalue $u_i$ which confines it around the origin. 
The right-hand side represents the sum of forces between $u_i$ and the others. 
The first term in the right-hand side provides a repulsive force, while the second term provides 
an attractive force. 
It is natural to expect that the values of $u_i$ are distributed around $u=0$, at least when $k$ is large enough so that the external Gaussian potential 
is dominant. 

The nature of the interaction among the eigenvalues depends of $n$. 
The case $n=0$ corresponds to pure Chern-Simons theory whose large $N$ solution is known \cite{Aganagic:2002wv}\cite{Halmagyi:2003ze}. 
There are long-range repulsive interactions, and due to them, the eigenvalue distribution can be extended largely when the 't Hooft coupling 
$\lambda:=N/k$ is large. 
For the case $n=1$, there is no long-range interactions, like in the case of ABJM theory. 
The large $N$ solution for this case was obtained in \cite{Suyama:2011yz}. 
In this case, the width of the eigenvalue distribution can be large for large $\lambda$, but its scaling with respect to $\lambda$ is different from 
neither the one in pure Chern-Simons theory nor in ABJM theory. 
For the remaining cases $n\ge2$, there are long-range attractive interactions. 
It it expected that the eigenvalues could not be largely extended in these cases. 

There is a quick way to guess whether the support of the eigenvalue distribution can be large if $k>0$. 
If $u_i$ is largely separated from most of the other eigenvalues, the hyperbolic functions in (\ref{SP0}) 
may be approximated by the sign function \cite{Suyama:2009pd}. 
The resulting equations 
\begin{equation}
\frac k{2\pi}u_i\ \sim\ (1-n)\sum_{j\ne i}\mbox{sgn}\left( \frac{u_i-u_j}2 \right)
\end{equation}
would make sense if $n<1$, while they look inconsistent if $n>1$. 
This suggests that the eigenvalues could be broadly distributed in the former case, but the eigenvalues would be always confined in a finite range 
in the latter cases. 

The level $k$ in (\ref{SP0}) must be replaced with $-ik$ with an integer $k$ so as to obtain some results relevant to ${\cal N}=3$ Chern-Simons-adjoint theories. 
This will be achieved via an analytic continuation of $k$ after the solution of (\ref{SP0}) for a positive real $k$ will be obtained. 

\vspace{5mm}

It is convenient to introduce new variables $z_i:=e^{u_i}$. 
In terms of them, the equations (\ref{SP0}) can be written as 
\begin{equation}
\log z_i\ =\ \frac tN\sum_{j\ne i}\frac{z_i+z_j}{z_i-z_j}-\frac{nt}N\sum_j\frac{z_i-z_j}{z_i+z_j}, 
   \label{SP}
\end{equation}
where $t:=2\pi \lambda$. 
The symmetry (\ref{symm-u}) implies 
\begin{equation}
\{\,z_1,\cdots,z_N\,\}\ =\ \{\,z_1^{-1},\cdots,z_N^{-1}\,\}. 
   \label{symm-z}
\end{equation}
The eigenvalues $\{z_i\}$ are distributed around $z=1$. 

%Note that 
%\begin{equation}
%\coth\frac{u_i-u_j}2\ =\ \frac{z_i+z_j}{z_i-z_j}, \hspace{5mm} \tanh\frac{u_i-u_j}2\ =\ \frac{z_i-z_j}{z_i+z_j}. 
%\end{equation}
%are used. 

A traditional strategy to solve (\ref{SP}) is to define a suitable complex function usually called the resolvent. 
A convenient choice of the resolvent in this case is 
\begin{equation}
v(z)\ :=\ \frac tN\sum_i\frac{z+z_i}{z-z_i}. 
   \label{resolvent-finite}
\end{equation}
For large $z$, $v(z)$ is expanded as 
\begin{equation}
v(z)\ =\ t+2t\langle W \rangle z^{-1}+O(z^{-2}), 
\end{equation}
where $\langle W \rangle$ is the vev of the Wilson loop (\ref{WL}). 
In the large $N$ limit, it is given as 
\begin{equation}
\langle W \rangle \ =\ \frac1N\sum_{i=1}^Nz_i. 
\end{equation}
The symmetry property (\ref{symm-z}) implies 
\begin{eqnarray}
v(z^{-1}) 
&=& \frac tN\sum_i\frac{z_i^{-1}+z}{z_i^{-1}-z} \nonumber \\
&=& -v(z), 
   \label{inversion}
\end{eqnarray}
where $z_i\ne0$ is assumed. 
This identity then implies that 
\begin{equation}
v(0)\ =\ -t
\end{equation}
which is compatible with the definition (\ref{resolvent-finite}). 

As usual in the matrix model literature, we assume that the large $N$ limit would be described by a continuous function $\rho(x)$ for the eigenvalue 
distribution. 
The meaning of $\rho(x)$ is 
\begin{equation}
\frac1N\sum_i\delta(x-z_i) \hspace{5mm} \stackrel{N\to\infty}{\longrightarrow} \hspace{5mm} \rho(x)
\end{equation}
in a suitable sense. 
The integral of $\rho(x)$ is normalized to be 1. 
According to the discussion above, the support of $\rho(x)$ consists of a single segment $[a,b]$ with 
\begin{equation}
ab\ =\ 1,
\end{equation}
which comes from the symmetry (\ref{symm-z}).  
The resolvent (\ref{resolvent-finite}) is written as 
\begin{equation}
v(z)\ =\ t\int dx\,\rho(x)\frac{z+x}{z-x}. 
   \label{resolvent-def}
\end{equation}
In terms of $v(z)$, the equations (\ref{SP}) can be written as 
\begin{eqnarray}
2\log y
%&=& t\cdot\frac1N\sum_{j\ne i}\frac{y+i0+z_j}{y+i0-z_j}+t\cdot\frac1N\sum_{j\ne i}\frac{y-i0+z_j}{y-i0-z_j}-2nt\cdot\frac1N\sum_j\frac{-y-z_j}{-y+z_j} 
%      \nonumber \\
&=& v(y+i0)+v(y-i0)-2n\,v(-y).  
   \label{SP2}
\end{eqnarray}
Here and in the following, the variable $y$ always takes a value in $[a,b]$. 

\vspace{5mm}

The expression (\ref{resolvent-def}) implies that $v(z)$ has the following analytic properties. 
\begin{itemize}
\item $v(z)$ is holomorphic on $\mathbb{C}\backslash[a,b]$. 
\item $v(z)$ is finite at $z=a,b$ since $\rho(x)$ vanishes at $x=a,b$. 
\item $v(z)$ is finite at infinity. 
\end{itemize}
It is shown in Appendix \ref{uniqueness} that the solution of (\ref{SP2}) with these analytic properties is unique.

\vspace{5mm}

\subsection{Ansatz for the resolvent} \label{section ansatz}

\vspace{5mm}

To solve the equation (\ref{SP2}), we make the following ansatz for $v(z)$. 
We assume that $v(z)$ can be written as 
\begin{equation}
v(z)\ =\ \int_{-\infty}^0 d\xi\,v(z,\xi), 
   \label{ansatz}
\end{equation}
where $v(z,\xi)$ is assumed to be a solution of 
\begin{equation}
-\frac2{\xi-1}\frac{y-1}{y-\xi}\ =\ v(y+i0,\xi)+v(y-i0,\xi)-2n\,v(-y,\xi). 
   \label{SP3}
\end{equation}
We further assume that $v(z,\xi)$ has the following analytic properties. 
\begin{itemize}
\item $v(z,\xi)$ is holomorphic on $\mathbb{C}\backslash[a,b]$. 
\item $v(z,\xi)$ is finite at $z=a,b$. 
\item $v(z,\xi)$ is finite at infinity. 
\end{itemize}
Note that $a,b$ are common for $v(z,\xi)$ with any $-\infty<\xi\le0$. 
The argument in Appendix \ref{uniqueness} shows that such a $v(z,\xi)$ is also unique. 

It is easy to show that $v(z)$ given as (\ref{ansatz}) satisfies (\ref{SP2}). 
Moreover, as long as $v(\infty)$ turns out to be finite, (\ref{ansatz}) has all the required analytic properties for $v(z)$. 
Therefore, due to the uniqueness of the solution of (\ref{SP2}), it turns out that one can obtain the solution of (\ref{SP2}) by solving (\ref{SP3}). 
The equation (\ref{SP3}) is a bit simpler than (\ref{SP2}) since the left-hand side of (\ref{SP2}) has a branch cut while that of (\ref{SP3}) has 
only a pole.

\vspace{1cm}

\section{${\cal N}=3$ Pure Chern-Simons theory} \label{chapter pureCS}

\vspace{5mm}

The simplest case $n=0$ corresponds to ${\cal N}=3$ pure Chern-Simons theory. 
The planar solution for this theory is known \cite{Aganagic:2002wv}\cite{Halmagyi:2003ze}. 
In this section, we show that our ansatz (\ref{ansatz}) can reproduce the known results for this theory. 

\subsection{Planar resolvent}

\vspace{5mm}

For the case $n=0$, the equation (\ref{SP3}) is simply 
\begin{equation}
-\frac2{\xi-1}\frac{y-1}{y-\xi}\ =\ v(y+i0,\xi)+v(y-i0,\xi). 
   \label{pureCS-SP}
\end{equation}

Define a function $\omega(z,\xi)$ such that $v(z,\xi)$ is written as 
\begin{equation}
v(z,\xi)\ =\ -\frac1{\xi-1}\frac{z-1}{z-\xi}+\omega(z,\xi). 
   \label{omega-def}
\end{equation}
Then, (\ref{pureCS-SP}) implies that $\omega(z,\xi)$ satisfies 
\begin{equation}
0\ =\ \omega(y+i0,\xi)+\omega(y-i0,\xi). 
\end{equation}
The general solution of this equation is of the form 
\begin{equation}
\omega(z,\xi) \ =\ A(z,\xi)\sqrt{(z-a)(z-b)}, 
\end{equation}
where $A(z,\xi)$ is a meromorphic function on $\mathbb{C}$. 
Note that the finiteness condition at $z=a,b$ was used. 

Since $v(z,\xi)$ is assumed to have no poles, $\omega(z,\xi)$ must have a pole at $z=\xi$ with the residue 1. 
This implies that $A(z,\xi)$ must have a pole at $z=\xi$. 
Therefore, $A(z,\xi)$ is of the form 
\begin{equation}
A(z,\xi)\ =\ \frac{A_1(z,\xi)}{z-\xi}, 
\end{equation}
where $A_1(z,\xi)$ is an entire function. 
Requiring the finiteness of $v(z,\xi)$ at infinty, $A_1(z,\xi)$ must be finite at infinity, implying that $A_1(z,\xi)$ is independent of $z$. 
The $\xi$-dependence of $A_1(z,\xi)$ is then determined by the required residue of the pole of $\omega(z,\xi)$ at $z=\xi$. 
It turns out to be 
\begin{equation}
A_1(z,\xi)\ =\ \frac1{\sqrt{(\xi-a)(\xi-b)}}. 
\end{equation}

It is concluded that $v(z,\xi)$ can be written as 
\begin{equation}
v(z,\xi)\ =\ -\frac1{\xi-1}\frac{z-1}{z-\xi}+\frac1{z-\xi}\frac{\sqrt{(z-a)(z-b)}}{\sqrt{(\xi-a)(\xi-b)}}. 
   \label{pureCS-soln}
\end{equation}
The resolvent $v(z)$ is obtained as the integral (\ref{ansatz}) of $v(z,\xi)$. 

The identity (\ref{inversion}) expected from the general argument can be verified explicitly. 
Using $ab=1$, one finds 
\begin{eqnarray}
v(z^{-1},\xi) 
%&=& -\frac1{\xi-1}\frac{1-z}{1-\xi z}+\frac1{1-\xi z}\frac{\sqrt{(1-az)(1-bz)}}{\sqrt{(\xi-a)(\xi-b)}} \nonumber \\
&=& \frac1{\xi^2}\left[ -\frac1{1-\xi^{-1}}\frac{1-z}{\xi^{-1}-z}+\frac1{\xi^{-1}-z}\frac{\sqrt{(1-az)(1-bz)}}{\sqrt{(1-a\xi^{-1})(1-b\xi^{-1})}} \right] \nonumber \\
&=& -\xi^{-2}v(z,\xi^{-1}), 
\end{eqnarray}
implying $v(z)=-v(z^{-1})$. 

\vspace{5mm}

Let us check whether the solution (\ref{pureCS-soln}) is the correct one. 
There is the following integral expression of the resolvent \cite{Suyama:2011yz}, 
\begin{eqnarray}
v(z) 
&=& 2\int_a^b\frac{dx}{2\pi}\frac{\log(x)}{z-x}\frac{\sqrt{(z-a)(z-b)}}{\sqrt{|(x-a)(x-b)|}}. 
   \label{old form}
\end{eqnarray}
Note that the definition of the resolvent in \cite{Suyama:2011yz} is slightly different from (\ref{resolvent-def}). 
Introducing a new variable 
\begin{equation}
x_\eta \ :=\  \frac{x-1}{(x-1)\eta+1}, 
\end{equation}
%Note that 
%\begin{eqnarray}
%\partial x_\eta &=& \frac1{[(x-1)\eta+1]^2}, \\
%dx 
%&=& [(x-1)\eta+1]^2dx_\eta, \nonumber \\
%&=& (\partial x_\eta)^{-1}dx_\eta, \\
%z-x 
%&=& [(z-1)\eta+1][(x-1)\eta+1](z_\eta-x_\eta) \nonumber \\
%&=& (\partial z_\eta)^{-\frac12}(\partial x_\eta)^{-\frac12}(z_\eta-x_\eta). 
%\end{eqnarray}
one can rewrite the integral in (\ref{old form}) as 
\begin{eqnarray}
2\int_a^b\frac{dx}{2\pi}\frac{\log x}{z-x}\frac{\sqrt{(z-a)(z-b)}}{\sqrt{|(x-a)(x-b)|}} 
&=& \int_0^1d\eta\oint_C\frac{dx}{2\pi i}\frac{x-1}{(x-1)\eta+1}\frac{1}{z-x}\frac{\sqrt{(z-a)(z-b)}}{\sqrt{(x-a)(x-b)}} \nonumber \\
%&=& \int_0^1d\eta\oint_{C_\eta}\frac{dx_\eta}{2\pi i}(\partial x_\eta)^{-1}\frac{x_\eta}{(\partial z_\eta)^{-\frac12}(\partial x_\eta)^{-\frac12}(z_\eta-x_\eta)} \nonumber \\
%& & \hspace*{1cm}\times \frac{(\partial z_\eta)^{-\frac12}\sqrt{(z_\eta-a_\eta)(z_\eta-b_\eta)}}{(\partial x_\eta)^{-\frac12}\sqrt{(x_\eta-a_\eta)(x_\eta-b_\eta)}} \nonumber \\
&=& \int_0^1d\eta\oint_{C_\eta}\frac{dx_\eta}{2\pi i}\frac{x_\eta}{z_\eta-x_\eta}\frac{\sqrt{(z_\eta-a_\eta)(z_\eta-b_\eta)}}{\sqrt{(x_\eta-a_\eta)(x_\eta-b_\eta)}}, 
\end{eqnarray}
where $C$ and $C_\eta$ are contours encircling the segment $[a,b]$ and $[a_\eta,b_\eta]$, respectively. 
Note that the following identity was used 
\begin{equation}
\log z\ =\ \int_0^1d\eta\,\frac{z-1}{(z-1)\eta+1}. 
   \label{another}
\end{equation}
The $x_\eta$-integral provides the resolvent of the Gaussian matrix model with the branch points $a_\eta$ and $b_\eta$. 
Using the explicit form of the resolvent of the Gaussian matrix model, one finds 
\begin{equation}
2\int_a^b\frac{dx}{2\pi}\frac{\log x}{z-x}\frac{\sqrt{(z-a)(z-b)}}{\sqrt{|(x-a)(x-b)|}} 
\ = \ \int_0^1d\eta\,\left[ z_\eta-\sqrt{(z_\eta-a_\eta)(z_\eta-b_\eta)} \right]. 
   \label{pureCS-check}
\end{equation}
By the change of integration variable $\eta=(1-\xi)^{-1}$, one can show that the integrand coincides with (\ref{pureCS-soln}). 
This confirms the validity of our ansatz (\ref{ansatz}) for the resolvent. 

\vspace{5mm}

Interestingly, the integral form (\ref{pureCS-check}) suggests that there is another way to obtain the solution of (\ref{SP2}) for the case $n=0$. 
Using the identity (\ref{another}), one finds that the solution of (\ref{SP2}) can be obtained from the solution of 
\begin{equation}
2y_\eta\ =\ v(y+i0,\eta)+v(y-i0,\eta). 
\end{equation}
Define $\tilde{v}(z,\eta)$ such that 
\begin{equation}
\tilde{v}(z_\eta,\eta)\ =\ v(z,\eta). 
\end{equation}
Then, $\tilde{v}(z,\eta)$ satisfies the saddle-point equation for the Gaussian matrix model with the branch cut on $[a_\eta,b_\eta]$. 
Due to the required analytic properties of $v(z,\eta)$, $\tilde{v}(z,\eta)$ must be holomorphic on $\mathbb{C}\backslash[a_\eta,b_\eta]$. 
In addition, $\tilde{v}(z,\eta)$ must be finite at infinite since 
\begin{equation}
z_\eta\ \to \ \infty \hspace{5mm} \Leftrightarrow \hspace{5mm} z\ \to\ -\frac{1-\eta}\eta. 
\end{equation}
These conditions imply that the resolvent is given as in (\ref{pureCS-check}). 

Unfortunately, this method cannot be applied to the general case $n>0$. 
If one writes down the saddle-point equation following the above strategy, one obtains 
\begin{equation}
2y_\eta\ =\ v(y_\eta+i0,\eta)+v(y_\eta-i0,\eta)-2n\,v((-y)_\eta,\eta). 
\end{equation}
Since $(-y)_\eta\ne -y_\eta$, one cannot simply borrow the results in, for example, \cite{Eynard:1995nv}\cite{Eynard:1995zv}.

\vspace{5mm}

\subsection{Analytic continuation}

\vspace{5mm}

The explicit formula (\ref{pureCS-soln}) for $v(z,\xi)$ and the relation $t=-v(0)$ provides an integral formula for the 't~Hooft coupling 
\begin{eqnarray}
t(a) 
&=& -\int_{-\infty}^0\frac{d\xi}{\xi}\left[ \frac1{\sqrt{(\xi-a)(\xi-b)}}+\frac1{1-\xi} \right]
   \label{t and a}
\end{eqnarray}
as a function of $a$. 
This integral is well-defined for $0<a\le1$. 
Note that $\sqrt{(\xi-a)(\xi-b)}$ is defined such that 
\begin{equation}
\sqrt{(\xi-a)(\xi-b)}\ \to\ \xi, \hspace{5mm} (\xi\to+\infty)
\end{equation}
so that the integrand is finite at $\xi=0$. 
The finiteness (\ref{t and a}) confirms the finiteness of  $v(z)$ at infinity which is necessary for the uniqueness of the solution of (\ref{SP2}). 

\begin{figure}[tbp]
\includegraphics{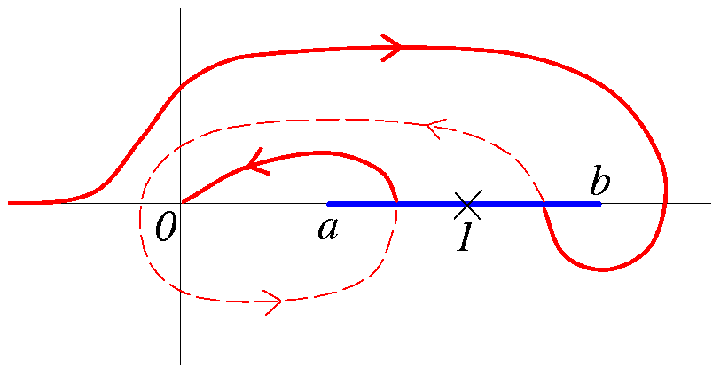}
\includegraphics{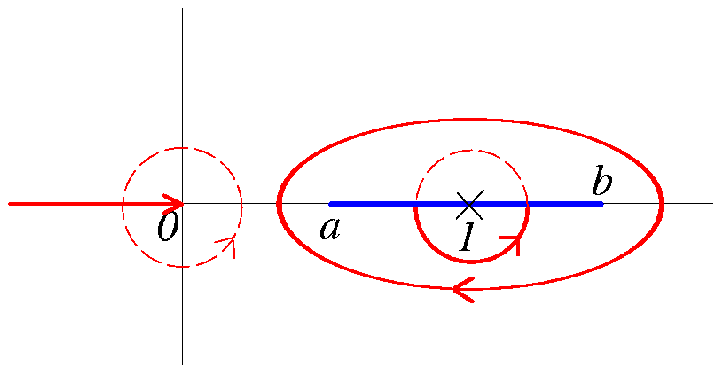}
\caption{
The deformed contour for the $2\pi$ rotation of the phase of $a$ (accompanied by $-2\pi$ rotation of the phase of $b$) (left). 
The contour on the right is an equivalent contour. 
The broken lines lie on the second sheet on which the integrand has a pole at the origin. 
The branch cut is shown by blue lines and the simple pole at $\xi=1$ is shown by the crosses.  
}
\label{contour}
\end{figure}

It is apparent from (\ref{t and a}) that $t(a)$ is always real for $0<a\le1$ and $t(1)=0$. 
An arbitrarily large $t(a)$ may be realized if there is a limit of $a$ such that the integral diverges. 
Indeed, $a\to0$ is such a limit, and the divergent contribution comes from the integration near $\xi=0$. 
One finds, for small $a$, 
\begin{eqnarray}
t(a) 
%&\sim& \int_0^1\frac{d\xi}\xi\left[ -\frac{\sqrt{a}}{\sqrt{\xi+a}}+1 \right] \nonumber \\
%&=& -\log a-2\log2+2\log(\sqrt{1+a}+\sqrt{a}) \nonumber \\
&=& -\log a+O(1). 
   \label{pureCS-small a}
\end{eqnarray}

To obtain some results relevant to pure Chern-Simons theory, an analytic continuation must be performed so that $t(a)$ becomes purely imaginary. 
This can be done by regarding $a$ as a complex variable, and then extend the domain of $t(a)$. 
Since the condition $ab=1$ is necessary for the integral (\ref{t and a}) to be convergent, the analytic continuation must be done while keeping this relation. 
Without loss of generality, one may assume $|a|\le|b|$. 
Therefore, $t(a)$ can be regarded as a function defined on a region 
\begin{equation}
D\ :=\ \{\,a\in\mathbb{C}\ |\ 0<|a|\le1, a\ne-1\,\}. 
\end{equation}
A point $a=-1$ was eliminated since for this value the two branch points $\xi=a,b$ are merged 
together on the integration contour, resulting in the divergence of $t(a)$. 
Since the integration contour is pinched by the branch points, there is no way to deform the contour to avoid the divergence. 

\begin{figure}[tbp]
\includegraphics{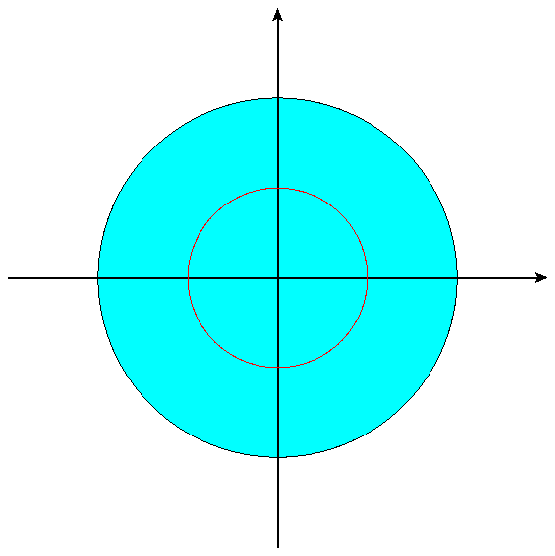}
\includegraphics{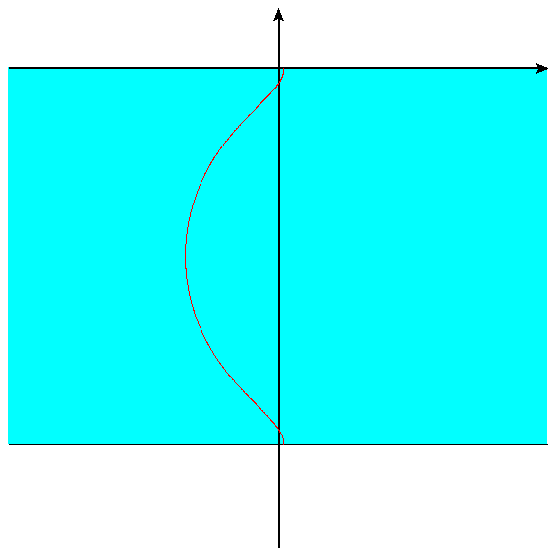}
\caption{
The unit disc in the $a$-plane (left) is mapped into an infinite strip with the width $2\pi$ in the $t$-plane (right). 
The red curves corresponds to $|a|=\frac12$. 
}
\label{pureCS-map}
\end{figure}

The function $t(a)$ is in fact a multi-valued function on $D$. 
Indeed, this can be anticipated from the log-behavior (\ref{pureCS-small a}). 
It turns out that the integral expression (\ref{t and a}) is very useful to analyze the analytic structure of $t(a)$. 
Consider the phase rotation of $a$. 
The branch points of the integrand move around in the complex $\xi$-plane, and due to this, the integration contour must be deformed to avoid 
the branch points. 
After the $2\pi$ rotation, the integration contour becomes the one depicted in Figure \ref{contour}. 
Figure \ref{contour} also shows an equivalent contour. 
Note that the integrand is smooth at $\xi=0$ on the first sheet, but has a simple pole with the residue 2 on the second sheet. 
Therefore, one obtains 
\begin{equation}
t(e^{2\pi i}a)\ =\ t(a)-2\pi i. 
   \label{pureCS-shift}
\end{equation}
This is consistent with the log-behavior (\ref{pureCS-small a}). 

As a map, $t(a)$ maps $D$ into an infinite strip in the complex $t$-plane, as depicted in Figure \ref{pureCS-map}. 
The only way to obtain a large imaginary part of $t(a)$ is to rotate the phase of $a$ many times. 
The modulus $|a|$ should be chosen to be of order $O(1)$ so as to make the real part of $t(a)$ vanish. 
As a result, the eigenvalue distribution, which is given by the branch cut of $v(z)$, has a length of order $O(1)$ even when $\mbox{Im}(t(a))$ is chosen 
to be large. 

\vspace{5mm}

All the results obtained above can be easily derived from the explicit relation \cite{Suyama:2011yz} 
\begin{equation}
\sqrt{a}+\frac1{\sqrt{a}}\ =\ 2e^{\frac 12t}. 
   \label{exact}
\end{equation}
It will turn out in subsection \ref{subsection CSA-t} that a similar analysis can be performed for the general cases $n\ge2$.

\vspace{5mm}

\subsection{Wilson loop}

\vspace{5mm}

As shown in subsection \ref{SPeq}, the vev $\langle W \rangle$ of the Wilson loop can be derived easily from the resolvent. 
One obtains an integral representation 
\begin{equation}
2t\langle W \rangle\ =\ \int_{-\infty}^0d\xi\left[ \frac{\xi-\frac{a+b}2}{\sqrt{(\xi-a)(\xi-b)}}-1 \right]. 
   \label{pureCS-WL}
\end{equation}
This integral is well-defined. 
Since the integrand has no pole on the whole Riemann sheet, the integral does not change by the $2\pi$ rotation of the 
phase of $a$. 
In other words, the integral defines a single-valued function on $D$. 
The integral diverges in the limit $a\to 0$ as 
\begin{equation}
2t\langle W \rangle\ =\ 2a^{-1}+O(1). 
\end{equation}
On the other hand, the integral is finite at $a=-1$. 

The vev $\langle W \rangle$ itself is a multi-valued function on $D$ since $t$ is multi-valued. 
For a fixed $|a|$, $|\langle W \rangle|$ decreases as the phase of $a$ decreases, or $\mbox{Im}(t)$ 
increases, since the integral (\ref{pureCS-WL}) is bounded on $|a|=$ const. 

One can check the validity of the above results by examining the explicit formula 
\begin{equation}
2t\langle W \rangle\ =\ \frac{(a-a^{-1})^2}{2(a^\frac12+a^{-\frac12})^2}. 
\end{equation}
Note that, using (\ref{exact}), one finds 
\begin{equation}
\langle W \rangle\ =\ e^{\frac 12t}\frac{\sinh\frac t2}{t}
\end{equation}
which is the large $N$ limit of the expression obtained in \cite{Kapustin:2009kz}.

\vspace{1cm}

\section{${\cal N}=3$ Chern-Simons-adjoint theories} \label{chapter CSA}

\vspace{5mm}

Let us return to the general case. 
The equation to be solved is 
\begin{equation}
-\frac2{\xi-1}\frac{y-1}{y-\xi}\ =\ v(y+i0,\xi)+v(y-i0,\xi)-2n\,v(-y,\xi). 
   \label{CSAdj-SP}
\end{equation}
In the following $n\ne1$ is assumed. 
The case $n=1$ was studied in \cite{Suyama:2011yz} based on a different technique. 

Let $f(z,\xi)$ be defined as 
\begin{equation}
f(z,\xi)\ :=\ \frac1{\xi-1}\frac{z-1}{z-\xi}. 
\end{equation}
As a function of $z$, $f(z,\xi)$ has a simple pole at $z=\xi$ with the residue $1$. 
Then (\ref{CSAdj-SP}) becomes 
\begin{equation}
-2f(y,\xi)\ =\ v(y+i0,\xi)+v(y-i0,\xi)-2n\,v(-y,\xi). 
\end{equation}
Define $\omega(z,\xi)$ such that $v(z,\xi)$ is written as 
\begin{equation}
v(z,\xi)\ =\ \frac{f(z,\xi)+nf(-z,\xi)}{n^2-1}+\omega(z,\xi). 
   \label{CSA-omega-def}
\end{equation}
The first term in the right-hand side is chosen such that 
\begin{equation}
2\frac{f(y,\xi)+nf(-y,\xi)}{n^2-1}-2n\frac{f(-y,\xi)+nf(y,\xi)}{n^2-1}\ =\ -2f(y,\xi). 
\end{equation}
Therefore, $\omega(z,\xi)$ satisfies the following homogeneous equation 
\begin{equation}
0 \ =\ \omega(y+i0,\xi)+\omega(y-i0,\xi)-2n\,\omega(-y,\xi). 
   \label{homogeneous}
\end{equation}

Fortunately, this equation was studied intensively in \cite{Eynard:1995nv}\cite{Eynard:1995zv} 
in the context of O$(N)$ matrix model \cite{Kostov:1988fy}\cite{Gaudin:1989vx}. 
In particular, an explicit solution in terms of the elliptic theta functions was found in \cite{Eynard:1995zv}.

\vspace{5mm}

\subsection{General solution of the homogeneous equation} \label{subsection homo}

\vspace{5mm}

We review the method for solving 
\begin{equation}
0 \ =\ \omega(y+i0)+\omega(y-i0)-2n\,\omega(-y)
   \label{homo}
\end{equation}
which was shown in \cite{Eynard:1995zv}. 
It is convenient to parametrize $n$ by $\nu$ as 
\begin{equation}
n\ =\ -\cos(\pi\nu). 
   \label{parametrizing n}
\end{equation}
Since our main interest is in the cases $n\ge2$, the parameter $\nu$ will be analytically continued to a complex value. 
Specifically, $\nu$ will be taken to be 
\begin{equation}
\nu\ =\ 1+i\tilde{\nu}, \hspace{5mm} \tilde{\nu}\ >\ 0. 
   \label{positive tilde-nu}
\end{equation}

The equation (\ref{homo}) can be written as 
\begin{equation}
\omega_{\pm}(y+i0)\ =\ -e^{\pm\pi i\nu}\omega_{\mp}(y-i0), 
   \label{homo3}
\end{equation}
where $\omega_\pm(z)$ are defined as 
\begin{equation}
\omega_{+}(z)\ :=\ \frac{e^{\frac12\pi i\nu}\omega(z)+e^{-\frac12{\pi i\nu}}\omega(-z)}{2\sin(\pi\nu)}, \hspace{5mm} 
 \omega_{-}(z)\ :=\ \omega_{+}(-z). 
   \label{omega_+-def}
\end{equation}

%Note that 
%\begin{eqnarray}
%0 
%&=& \omega(y+i0)+\omega(y-i0)+(e^{\pi i\nu}+e^{-\pi i\nu})\omega(-y) \nonumber \\
%&=& e^{-\frac{\pi i\nu}2}\left[ e^{\frac{\pi i\nu}2}\omega(y+i0)+e^{-\frac{\pi i\nu}2}\omega(-y) \right]
%      +e^{\frac{\pi i\nu}2}\left[ e^{-\frac{\pi i\nu}2}\omega(y-i0)+e^{\frac{\pi i\nu}2}\omega(-y) \right] \nonumber \\
%&=& e^{-\frac{\pi i\nu}2}\left[ e^{\frac{\pi i\nu}2}\omega(y-i0)+e^{-\frac{\pi i\nu}2}\omega(-y) \right]
%      +e^{\frac{\pi i\nu}2}\left[ e^{-\frac{\pi i\nu}2}\omega(y+i0)+e^{\frac{\pi i\nu}2}\omega(-y) \right]. 
%   \label{homo2}
%\end{eqnarray}

There exists a pair of functions $G_\pm(z)$ which satisfy (\ref{homo3}) \cite{Eynard:1995nv}\cite{Eynard:1995zv}. 
The properties of $G_\pm(z)$ are summarized in Appendix \ref{G(z)}. 
The general solution of (\ref{homo3}) is obtained in terms of $G_\pm(z)$. 
Let $S_\pm(z)$ be defined as 
\begin{equation}
S_\pm(z)\ :=\ \frac{\omega_{\pm}(z)}{G_\pm(z)}. 
\end{equation}
Then, $S_\pm(z)$ satisfy 
\begin{equation}
S_\pm(y+i0)\ =\ S_\mp(y-i0). 
\end{equation}
The implications of these equations become transparent when they are written in terms of the following functions, 
\begin{eqnarray}
S(z) &:=& S_+(z)+S_-(z)\ =\ S_+(z)+S_+(-z), \\
\tilde{S}(z) &:=& S_+(z)-S_-(z)\ =\ S_+(z)-S_+(-z). 
\end{eqnarray}
Note that $S(z)$ is an even function while $\tilde{S}(z)$ is odd. 
They satisfy 
\begin{equation}
S(y+i0)\ =\ S(y-i0), \hspace{5mm} \tilde{S}(y+i0)\ =\ -\tilde{S}(y-i0). 
\end{equation}
These equations imply 
\begin{equation}
S(z)\ =\ 2A(z^2), \hspace{5mm} \tilde{S}(z)\ =\ 2zB(z^2)\sqrt{(z^2-a^2)(z^2-b^2)}, 
\end{equation}
where $A(z^2)$ and $B(z^2)$ are meromorphic functions on $\mathbb{C}$. 
In terms of these functions, $\omega_\pm(z)$ can be written as 
\begin{equation}
\omega_{\pm}(z)\ =\ \left[ A(z^2)\pm zB(z^2)\sqrt{(z^2-a^2)(z^2-b^2)} \right]G_\pm(z). 
\end{equation}
Since $\omega_-(z)$ is obtained as $\omega_+(-z)$, it is enough to determine $\omega_+(z)$. 

\vspace{5mm}

As reviewed in Appendix \ref{G(z)}, $G_+(z)$ has singularities at $z=a,b$ where $G_+(z)$ 
behaves as $(z-a)^{-\frac12}$ and $(z-b)^{-\frac12}$, respectively. 
Assuming that $\omega_+(z)$ is finite at these points, it must be of the form 
\begin{equation}
\omega_+(z)\ =\ \left[ A_1(z^2)\sqrt{(z^2-a^2)(z^2-b^2)}+zB(z^2) \right]\sqrt{(z^2-a^2)(z^2-b^2)}G_+(z). 
\end{equation}

$G_+(z)$ has a simple zero at $z=e$ ($e$ is determined by $a$ and $b$. See Appendix \ref{G(z)}). 
This allows both $A_1(z^2)$ and $B(z^2)$ to have simple poles at $z=e$ 
even when $\omega_+(z)$ is assumed to be regular at $z=e$. 
Then, one can write $\omega_+(z)$ as 
\begin{equation}
\omega_{+}(z)\ =\ \left[ A_2(z^2)\frac{\sqrt{(z^2-a^2)(z^2-b^2)}}{z^2-e^2}+z\frac{B_1(z^2)}{z^2-e^2} \right]
 \sqrt{(z^2-a^2)(z^2-b^2)}G_+(z). 
\end{equation}
Let us assume further that $\omega_+(z)$ is regular at $z=-e$. 
Then, the pole at $z=-e$ in the above expression must be canceled. 
Rewriting $B_1(z^2)$ as 
\begin{equation}
B_1(z^2)\ =\ \frac{\bar{e}}eA_2(z^2)+B_2(z^2), \hspace{5mm} \bar{e}\ :=\ \sqrt{(e^2-a^2)(e^2-b^2)}, 
\end{equation}
one obtains 
\begin{eqnarray}
\omega_{+}(z)
%&=& \left[ A_2(z^2)\left( \frac{\sqrt{(z^2-a^2)(z^2-b^2)}+\frac{\bar{e}}ez}{z^2-e^2} \right)+z\frac{B_2(z^2)}{z^2-e^2} \right]\sqrt{(z^2-a^2)(z^2-b^2)}G_+(z)
%       \nonumber \\
&=& \left[ A_2(z^2)g(z)+z\frac{B_2(z^2)}{z^2-e^2} \right]\sqrt{(z^2-a^2)(z^2-b^2)}G_+(z), 
\end{eqnarray}
where 
\begin{equation}
g(z) \ :=\ \frac{\sqrt{(z^2-a^2)(z^2-b^2)}+\frac{\bar{e}}ez}{z^2-e^2}. 
   \label{g(z)-def}
\end{equation}
The regularity at $z=-e$ then implies $B_2(z^2)=(z^2-e^2)B_3(z^2)$. 

\vspace{5mm}

In summary, it has been shown that the solution of the equation (\ref{homo3}) which is finite at $z=a,b$ and regular at $z=\pm e$ has the general form 
\begin{equation}
\omega_{+}(z)\ =\ \left[ A(z^2)g(z)+zB(z^2) \right]\sqrt{(z^2-a^2)(z^2-b^2)}G_+(z), 
   \label{general soln}
\end{equation}
where $A(z^2)$ and $B(z^2)$ are meromorphic functions regular at $z=a,b,\pm e$.

\vspace{5mm}

\subsection{Planar resolvent} \label{subection CSA-res}

\vspace{5mm}

The function $\omega(z,\xi)$ defined in (\ref{CSA-omega-def}) depends on the parameter $\xi$. 
Recall that the parameters $a,b$ are assumed to be independent of $\xi$. 
Therefore, the solution in the previous section can be used for $\omega_+(z,\xi)$ simply by allowing the functions $A(z^2)$ and $B(z^2)$ to depend on 
$\xi$. 
That is, $\omega_+(z,\xi)$ can be written as  
\begin{equation}
\omega_+(z,\xi)\ =\ \left[ A(z^2,\xi)g(z)+zB(z^2,\xi) \right]\sqrt{(z^2-a^2)(z^2-b^2)}G_+(z). 
\end{equation}
It will be shown below that $A(z^2,\xi)$ and $B(z^2,\xi)$ are determined by the analytic properties of $\omega_+(z,\xi)$ expected from 
those of $v(z,\xi)$ assumed in subsection \ref{section ansatz}. 

\vspace{5mm}

The definition (\ref{CSA-omega-def}) of $\omega(z,\xi)$ implies that $\omega_+(z,\xi)$ is related to $v(z,\xi)$ as 
\begin{eqnarray}
\omega_{+}(z,\xi) 
&=& v_{+}(z,\xi)-\frac {i}{2(n^2-1)}\left( e^{-\frac12{\pi i\nu}}f(z,\xi)-e^{\frac12{\pi i\nu}}f(-z,\xi) \right), 
   \label{omega and v}
\end{eqnarray}
where $v_+(z,\xi)$ is defined from $v(z,\xi)$ as in (\ref{omega_+-def}). 
This relation then implies that $\omega_+(z,\xi)$ is finite at infinity. 
This also implies that $\omega_+(z,\xi)$ must have simple poles at $z=\pm\xi$ with the residues 
\begin{eqnarray}
\mbox{Res}_\xi \ \omega_+(z,\xi) &=& -\frac i{2(n^2-1)}e^{-\frac12{\pi i\nu}}, 
   \label{res1} \\
\mbox{Res}_{-\xi} \ \omega_+(z,\xi) &=& -\frac i{2(n^2-1)}e^{\frac12{\pi i\nu}}. 
   \label{res2}
\end{eqnarray}
These poles must come from $A(z^2,\xi)$ and $B(z^2,\xi)$. 
Therefore, they can be written as 
\begin{equation}
A(z^2,\xi)\ =\ \frac{A_1(z^2,\xi)}{z^2-\xi^2}, \hspace{5mm} B(z^2,\xi)\ =\ \frac{B_1(z^2,\xi)}{z^2-\xi^2}, 
\end{equation}
where $A_1(z^2,\xi)$ and $B_1(z^2,\xi)$ are entire functions of $z$. 
The behavior of $\omega_+(z,\xi)$ near infinity is then 
\begin{eqnarray}
\omega_{+}(z,\xi) 
%&\sim& \left[ z^{-2}A_1(z^2,\xi)+z^{-1}B_1(z^2,\xi) \right]iz \nonumber \\
&\sim& iz^{-1}A_1(z^2,\xi)+iB_1(z^2,\xi), 
\end{eqnarray}
where 
\begin{equation}
G_+(z)\ =\ \frac iz+O(z^{-2})
\end{equation}
was used. 
See Appendix \ref{G(z)}. 
The finiteness of $\omega_+(z,\xi)$ at infinity implies 
\begin{equation}
A_1(z^2,\xi)\ =\ c(\xi), \hspace{5mm} B_1(z^2,\xi)\ =\ d(\xi). 
\end{equation}
The constants $c(\xi)$ and $d(\xi)$ are then determined by requiring that $\omega_+(z,\xi)$ has the correct residues (\ref{res1})(\ref{res2}) at $z=\pm\xi$. 
One finds 
\begin{equation}
c(\xi)\ =\ -\frac1{2(n^2-1)}\xi G(\xi), \hspace{5mm} d(\xi)\ =\ \frac{1}{2(n^2-1)e}\xi^{-1}G(\xi^{-1}), 
\end{equation}
where the following identity was used, 
\begin{equation}
e\,g(z)G_+(z)\ =\ z^{-1}G_+(z^{-1}), 
\end{equation}
which is valid when $ab=1$ is satisfied. 
See Appendix \ref{identity} and \ref{coeff} for the details. 

\vspace{5mm}

Now, $\omega_+(z,\xi)$ has been determined completely. 
It can be written in the following form 
\begin{equation}
\omega_+(z,\xi)\ =\ -\frac 1{2(n^2-1)e}\Bigl[ \tilde{\omega}(z,\xi)-\xi^{-2}\tilde{\omega}(z^{-1},\xi^{-1}) \Bigr], 
   \label{CSA-res-final}
\end{equation}
where 
\begin{equation}
\tilde{\omega}(z,\xi)\ :=\ \frac{\sqrt{(z^2-a^2)(z^2-b^2)}}{z^2-\xi^2}\xi G(\xi)z^{-1}G_+(z^{-1}). 
\end{equation}

It is straightforward to verify the identity (\ref{inversion}) by noticing 
\begin{eqnarray}
f(z^{-1},\xi) &=& -\xi^{-2}f(z,\xi^{-1}), \\
\omega_+(z^{-1},\xi) &=& -\xi^{-2}\omega_+(z,\xi^{-1}). 
\end{eqnarray}

\vspace{5mm}

One might be worried that the results obtained above would depend on a particular choice of $\nu$. 
Indeed, since $n=1$ is excluded in our calculation, there are at least two ways to continue $n$ from 0, which was studied in section 
\ref{chapter pureCS}, to the cases $n\ge2$. 
Understanding (\ref{parametrizing n}) as a map from the complex $n$-plane to the complex $\nu$-plane, the upper-half $n$-plane is mapped into 
\begin{equation}
R_+\ :=\ \{\,\nu\in\mathbb{C}\ |\ 0<\mbox{Re}(\nu)<1,\ \mbox{Im}(\nu)>0\,\}, 
\end{equation}
and the lower-half $n$-plane is mapped into 
\begin{equation}
R_-\ :=\ \{\,\nu\in\mathbb{C}\ |\ 0<\mbox{Re}(\nu)<1,\ \mbox{Im}(\nu)<0\,\}. 
\end{equation}
These two choices correspond to taking $\tilde{\nu}$ to be positive, as in (\ref{positive tilde-nu}), or to be negative. 
It can be checked that both prescriptions for the analytic continuation 
provide the same $\omega(z)$ by noticing that the only change due to the change of the prescription 
is the sign flip of $G(z)$. 
The independence of the prescription is also a consequence of the uniqueness of the solution of (\ref{homo}) based on an argument similar to Appendix 
\ref{uniqueness}.

\vspace{5mm}

\subsection{Analytic continuation} \label{subsection CSA-t}

\vspace{5mm}

\begin{figure}[tbp]
\includegraphics{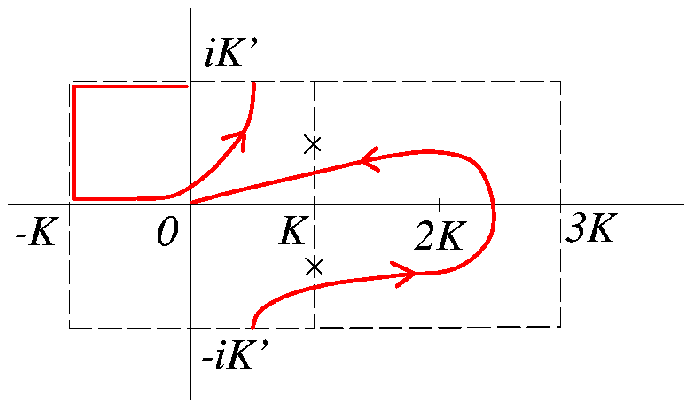}
\includegraphics{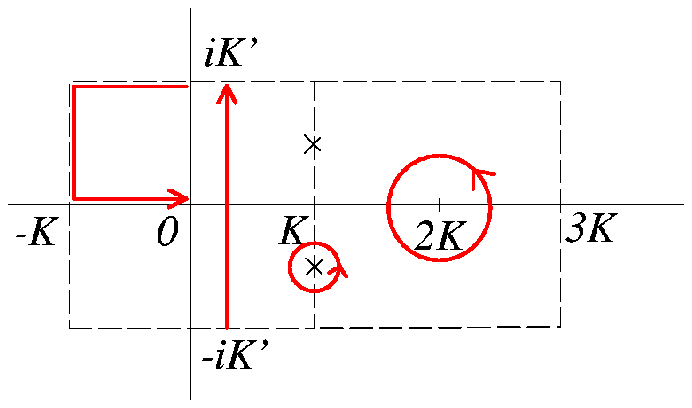}
\caption{
The contour in Figure \ref{contour} is drawn in the $u$-plane (left). 
The upper and the lower edges of the rectangles are identified. 
The crosses correspond to $\xi=1$. 
An equivalent contour is also drawn (right). 
}
\label{u-contour}
\end{figure}

Recall that the 't Hooft coupling $t$ is obtained as a function of $a$ through the resolvent as 
\begin{eqnarray}
t(a)
&=& -v(0) \nonumber \\
&=& -\int_{-\infty}^0d\xi\,\left[ \omega(0,\xi)+\frac1{n-1}\frac1{\xi(\xi-1)} \right]. 
\end{eqnarray}
The explicit expression of (\ref{CSA-res-final}) gives 
\begin{equation}
t(a)\ =\ \frac1{n-1}\int_{-\infty}^0\frac{d\xi}\xi\ \left[ \frac i{2e\sin\frac{\pi\nu}2}G(\xi)+\frac1{1-\xi} \right]. 
   \label{tHooft for CSA}
\end{equation}
Note that this integral expression is quite similar to (\ref{t and a}) for pure Chern-Simons theory. 

This integral is well-defined for $0<a\le1$. 
Indeed, for large $\xi$, the integrand behaves as $\xi^{-2}$, while at $\xi=0$ the function in the square brackets vanishes since 
\begin{equation}
G(0)\ =\ 2\sin\frac{\pi\nu}2G_+(0)\ =\ 2ie\sin\frac{\pi\nu}2. 
\end{equation}
The finiteness also implies that $v(\infty)$ is finite, as expected. 

The integral (\ref{tHooft for CSA}) defines a function on $D$. 
As in the case of pure Chern-Simons theory, $t(a)$ is a multi-valued function. 
One may expect this property from the fact that $G(z)$ has the branch cut on $[a,b]$, the same position as $((z-a)(z-b))^{-\frac12}$. 
Due to this analytic structure, 
the effect of the phase rotation of $a$ by $2\pi$ is determined by the deformation of the integration contour which is exactly the same one 
depicted in Figure \ref{contour}. 
It is more convenient to consider the contour in the $u$-plane. 
The deformed contour in the $u$-plane is shown in Figure \ref{u-contour}. 
Note that the vertical contour in Figure \ref{u-contour} is equivalent to a contour encircling the segment $[-b,-a]$ in the $\xi$-plane. 
Since the integrand has no branch cut on $[-b,-a]$, the contribution from this contour vanishes. 
Now, the contour deformation amounts to adding residues of two poles, $\xi=1$ and $\xi=0$ on the second sheet ($u=2K$). 
One finds 
\begin{eqnarray}
t(e^{2\pi i}a)
&=& t(a)+\frac{2\pi i}{n-1}\left[ \frac{\sin\frac{3\pi\nu}2}{\sin\frac{\pi\nu}2}+1-1 \right] \nonumber \\
&=& t(a)-2\pi i\,\frac{2n-1}{n-1}. 
\end{eqnarray}
For the case $n=0$, this formula coincides with (\ref{pureCS-shift}) for pure Chern-Simons theory. 
The singularity at $n=1$ seems to be compatible with the results in \cite{Suyama:2011yz}. 

\begin{figure}[tbp]
\begin{center}
\includegraphics{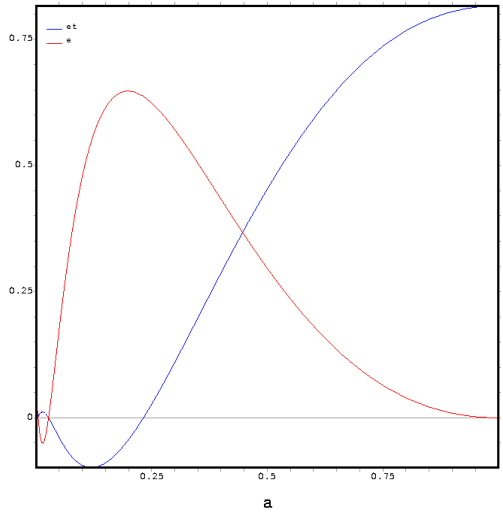}
\end{center}
\caption{
The red curve is the plot of $e\,t(a)$ against $a$, and the blue curve is the plot of $e$ against $a$. 
$n=5$ is chosen. 
}
\label{plot1}
\end{figure}

A plot of $e\, t(a)$ for real $a$ with a fixed $n$ is shown in Figure \ref{plot1}. 
The choice $e\,t(a)$ is due to the fact that a factor $e^{-1}$ exists in the integrand which may diverge for some $a$. 
Figure \ref{plot1} shows that $t(a)$ increases from 0 as $a$ decreases from 1. 
The positivity of $t(a)$ for a real $a$ can be regarded as a consistency check of our resolvent since it is required from the definition. 
Note that $e$ is also a function of $a$ which may vanish even when $e\,t(a)$ is non-zero. 
This indicates that $t(a)$ diverges to $+\infty$ at $a=a_*$ where $a_*$ satisfies 
\begin{equation}
\tilde{\nu}K'\ =\ 2K. 
   \label{zero of e}
\end{equation}
See Appendix \ref{G(z)} for the dependence of $K,K'$ on $a$. 
In other words, $t(a)$ has a simple pole at $a=a_*$. 
Recalling that $e$ is proportional to the elliptic function sn$(\tilde{\nu}K',k)$ and that $K'$ increases as $a$ decreases, 
one finds that $t(a)$ has an infinite number of simple poles on the real axis in $D$. 

If $a$ becomes smaller than $a_*$, then $t(a)$ becomes negative. 
This is certainly not a signal of the inconsistency of the resolvent, but the indication that for any real and positive 't Hooft coupling, $a$ is always 
within the range $(a_*,1]$. 
In other words, the width of the eigenvalue distribution is always of order $O(1)$. 
This behavior of the eigenvalue distribution fits nicely with the intuition obtained from the original saddle-point equations (\ref{SP0}). 
For the case $n>1$, there are long-range attractive interactions among eigenvalues, so they would be bound within a finite region. 
On the other hand, one can check that $e$ never vanishes for the case $n<1$. 
Due to this fact, $a$ can be arbitrarily close to 0 before $t(a)$ diverges, resulting in a largely extended eigenvalue distribution. 
Correspondingly, there are long-range repulsive interactions among eigenvalues for the case $n<1$. 

\vspace{5mm}

\begin{figure}[tbp]
\begin{center}
\includegraphics{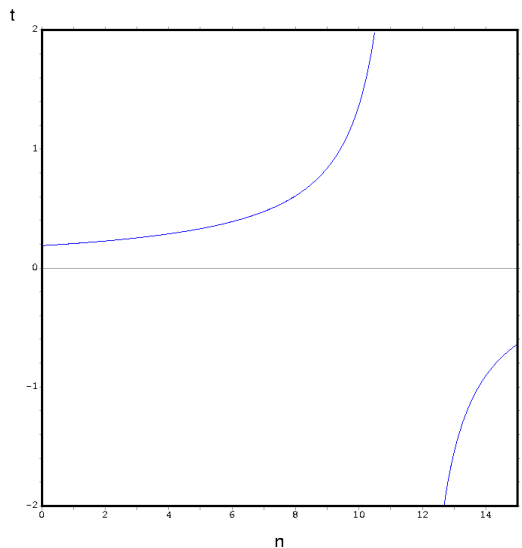}
\end{center}
\caption{
$t(a)$ is plotted against $n$ with $a$ chosen such that ${\frac{K'}{K}}=2$. 
$t(a)$ diverges at $n=\cosh\pi\sim 11.6$ according to (\ref{zero of e}). 
}
\label{plot2}
\end{figure}

It would be interesting to see the dependence of $t(a)$ on $n$. 
Figure \ref{plot2} shows a plot of $t(a)$ as a function of $n$ with a fixed $a$. 
The value of $t(a)$ for $n=0$ reproduces the known exact value 
\begin{equation}
t_{n=0}(a)\ =\ 2\log\left( \sqrt{a}+\frac1{\sqrt{a}} \right)-2\log2. 
\end{equation}
It turns out that $t(a)$ is positive as $n$ increases from 0 up to some value. 
Interestingly, $t(a)$ is smooth at $n=1$ although the expression (\ref{tHooft for CSA}) looks singular. 
As $n$ increases further, Figure \ref{plot2} shows that $t(a)$ diverges to $+\infty$. 
This is again due to the presence of the factor $e^{-1}$ in (\ref{tHooft for CSA}). 

\vspace{5mm}

\begin{figure}[tbp]
\includegraphics{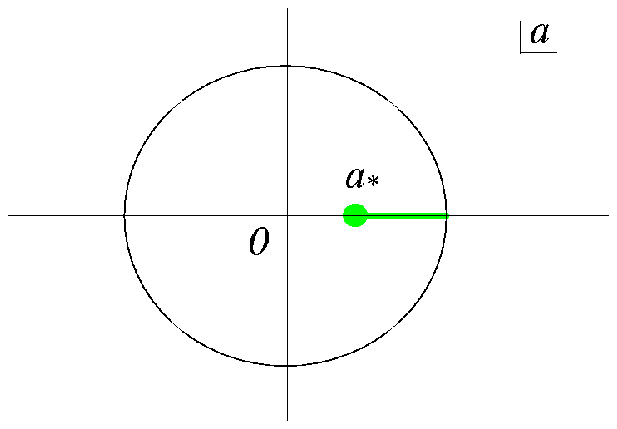}
\includegraphics{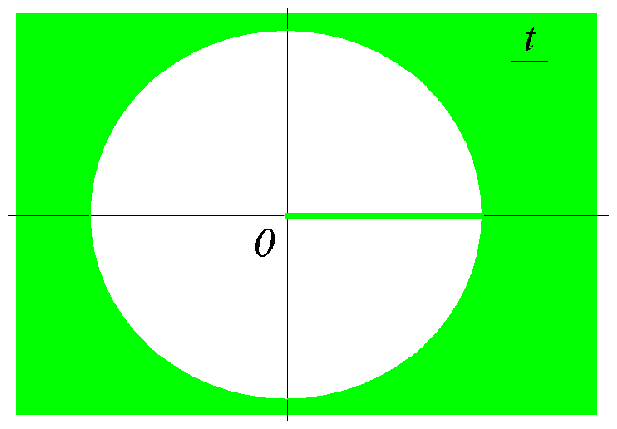}
\caption{
A small disc centered at $a=a_*$ in the $a$-plane is mapped to a neighborhood of infinity in the $t$-plane. 
}
\label{map2}
\end{figure}

The issue of the analytic continuation from a real $t(a)$ to $t(a)=2\pi i\lambda$ with a large positive $\lambda$ is rather subtle. 
For pure Chern-Simons theory, there is a unique way to obtain such a value of $t(a)$. 
This is because, although $t(a)$ is multi-valued, its inverse function $a(t)$ is single-valued. 
However, for the case $n>1$, there are infinitely many values of $a$ for which Im$(t(a))$ becomes large. 
One way to obtain a large Im$(t(a))$ is, as for pure Chern-Simons theory, to rotate the phase of $a$ many times to increase Im$(t(a))$. 
Another way is to choose one pole of $t(a)$, say $a=a_*$, and choose one point in the vicinity of $a_*$. 
Because of the pole at $a=a_*$, $t(a)$ maps a small disc centered at $a_*$ in the $a$-plane to a neighborhood of infinity in the $t$-plane. 
This is depicted in Figure \ref{map2}. 
Therefore, by giving a tiny imaginary part to $a$ near $a_*$, one obtains a large and positive $\lambda$. 
If the latter continuation would be chosen, then observables would be insensitive to the analytic continuation as long as the observables are 
continuous at $a=a_*$. 

The ambiguity for the prescription of the analytic continuation can be also described as follows. 
The function $t(a)$ has many poles on its Riemann sheet, which consists of an infinite number of $D$ glued together, 
and also diverges along the direction corresponding to the infinite phase rotation, 
Therefore, a given large value of Im$(t(a))$ does not uniquely specify the value of $a$, and therefore, the planar resolvent. 
One of the most natural choice would be to choose the pole $a=a_*$ since it is directly connected to $a\sim 1$ or $t(a)\sim 0$ for which 
the matrix model is apparently well-defined. 

Note that the width of the eigenvalue distribution is of order $O(1)$ independent of the prescription of the analytic continuation. 

\vspace{5mm}

\subsection{Wilson loop}

\vspace{5mm}

An integral formula for the vev $\langle W \rangle$ of the Wilson loop is 
\begin{equation}
2t\langle W \rangle\ =\ 
 -\frac1{n+1}\int_{-\infty}^0d\xi\left[ \frac{\cos\frac{\pi\nu}2}{n-1}\left( \xi G(\xi)-\frac Ce\xi^{-1}G(\xi^{-1}) \right)+1 \right], 
   \label{CSA-WL}
\end{equation}
where $C$ is a constant appearing in the large $z$ expansion of $G_+(z)$, 
\begin{equation}
G_+(z)\ =\ \frac iz+\frac{iC}{z^2}+O(z^{-3}). 
\end{equation}
The integral is well-defined. 
Note that the integral decreases fast enough for large $\xi$ since 
\begin{equation}
\xi G(\xi)-\frac Ce\xi^{-1}G(\xi^{-1})\ =\ 2\cos\frac{\pi\nu}2+O(\xi^{-2}). 
\end{equation}
Since there is no pole in the integrand, the integral defines a single-valued function on $D$. 
Note that $2t\langle W \rangle$ may diverge at some $a$ due to the factor $e^{-1}$, 
but this divergence is due to the divergence of $t$, and therefore $\langle W \rangle$ 
is finite for a non-zero $a$.

The large $\lambda$ behavior may depend on the choice of the analytic continuation. 
If one obtains a large imaginary part of $t$ by rotating the phase of $a$ many times, then $\langle W \rangle$ decreases as $O(t^{-1})$ 
since the integral (\ref{CSA-WL}) is bounded for a fixed $|a|$. 
One the other hand, if one choose an analytic continuation in which $a$ is chosen to be $a_*$ plus a suitable infinitesimal imaginary part, 
then one finds $\langle W \rangle=O(1)$.

\vspace{1cm}

\section{Discussion} \label{discuss}

\vspace{5mm}

In this paper, we determined the planar resolvent of ${\cal N}=3$ Chern-Simons-adjoint theories. 
The resolvent was given as an integral of a combination of elliptic theta functions. 
Although the expression looks complicated, it is explicit enough to investigate its analytic structure. 
It was clarified that there are various ways to continue the 't Hooft coupling $t$ to a purely imaginary value by which one can hope to extract some 
information on Chern-Simons-adjoint theories. 
For those analytic continuations, the width of the eigenvalue distribution does not grow indefinitely in the large $\mbox{Im}(t)$ limit 
which can be deduced from the saddle-point equations (\ref{SP0}). 
Note that a similar behavior was observed in \cite{Marino:2012az} for an ${\cal N}=2$ version of Chern-Simons-adjoint theories. 
Accordingly, the vev $\langle W \rangle$ of a BPS Wilson loop is finite. 
It was observed that different behaviors of $\langle W \rangle$ may result from different analytic continuations. 

It is interesting to generalize the analysis performed in this paper to more general Chern-Simons-matter theories. 
It is expected that the addition of fundamental matters would be rather straightforward since it turned out to be the case for pure Chern-Simons theory 
\cite{Suyama:2010hr} and for ABJM theory \cite{Santamaria:2010dm}. 
Inclusion of matters in the symmetric and the anti-symmetric representations would be possible in the following way. 
A pair of the symmetric matters would contribute to the saddle-point equations (\ref{SP0}) as the following additional terms 
\begin{equation}
-\frac12\sum_{j\ne i}\tanh\frac{u_i+u_j}2-\tanh u_i. 
\end{equation}
Note that these terms does not spoil the symmetry (\ref{symm-u}). 
In terms of $z_i$, these terms can be written as 
\begin{eqnarray}
 -\frac12\sum_{j}\frac{z_i-z_j^{-1}}{z_i+z_j^{-1}}-\frac12\frac{z_i-z_i^{-1}}{z_i+z_i^{-1}} 
&=& -\frac12\sum_{j}\frac{z_i-z_j}{z_i+z_j}-\frac12\frac{z_i-z_i^{-1}}{z_i+z_i^{-1}}, 
\end{eqnarray}
where the symmetry (\ref{symm-z}) was used. 
Therefore, the effects of adding a pair of the symmetric matters appear as a shift of $n$ by $\frac12$ in addition to a modification of the external force 
acting on $u_i$ or $z_i$. 
The effects of anti-symmetric matters is almost the same, but the sign of the extra external force is opposite. 

It would be more interesting to investigate Chern-Simons-matter theories with multiple gauge group factors. 
The analysis in this paper would be able to be generalized to quiver-type theories with two nodes. 
In these theories, one can also consider bi-fundamental matters. 
A systematic analysis of such a family of Chern-Simons-matter theories would provide an interesting pattern in the eigenvalue distribution and the 
behavior of the Wilson loop. 

Of course, it would be interesting to ask whether there exists a dual description for those Chern-Simons-matter theories. 
Since the Wilson loop in most of the theories mentioned above would not exhibit the exponential behavior, which would be expected if 
there exists a dual classical worldsheet in AdS space, the gravity dual would not be simple even if it exists. 
The possible dual theory might be a higher spin theory, as in \cite{Minwalla:2011ma}\cite{Chang:2012kt}\cite{Aharony:2012nh}, 
or it might be something else. 
The detailed study of the planar solution may provide some information necessary for this line of research. 

Very recently, it was shown in \cite{Kallen:2012zn} 
that the free energy of ${\cal N}=2$ super Yang-Mills theory in five dimensions scales as $N^3$, as expected from the 
conjectured relation to (2,0)-theory in six dimensions \cite{Douglas:2010iu}\cite{Lambert:2010iw}. 
Curiously, the same calculation can be performed for ${\cal N}=3$ pure Chern-Simons theory with {\it purely imaginary} level, and the free energy scales 
as $N^3$. 
Note that this scaling comes as $\lambda N^2$. 
In addition, it was shown in \cite{Kim:2012av} that a modified partition function has a localization formula whose perturbative part coincides with the one for 
pure Chern-Simons theory. 
In pure Chern-Simons theory, since the level must be an integer, the above leading term becomes the imaginary part of the free energy. 
This observation might shed some light on the physical meaning of the $N^3$ scaling behavior.

\vspace{2cm}

{\bf \Large Acknowledgements}

\vspace{5mm}

I would like to thank Jean-Emile Bourgine and Soo-Jong Rey for valuable comments. 
This work was supported in part by the BK21 program of the Ministry of Education, Science and Technology, 
National Science Foundation of Korea Grants 0409-20110151, 2005-009-3843, 2009-008-0372 and 2010-220-C00003.

\newpage

\appendix

\vspace{1cm}

\section{$G(z)$} \label{G(z)}

\vspace{5mm}

This appendix shows some properties of a function $G(z)$ by reviewing the derivation given in \cite{Eynard:1995zv}. 

The function $G(z)$ is a solution of the homogeneous equation 
\begin{equation}
G(y+i0)+G(y-i0)-2n\,G(-y)\ =\ 0. 
   \label{homo-appendix}
\end{equation}
It is assumed that $G(z)$ is a holomorphic function on $\mathbb{C}\backslash[a,b]$ and that $G(z)$ has a branch cut on $[a,b]$. 
Due to this branch cut, the Riemann sheet of $G(z)$ would be non-trivial. 
To make the analysis of the Riemann sheet tractable, it is convenient to introduce a new variable $u$ via the Jacobi's elliptic function, 
\begin{equation}
z\ = \ a\,\mbox{sn}(u,k), 
   \label{u-def}
\end{equation}
where $k:=\displaystyle{\frac ab}$. 
Note that $k$ in this appendix has nothing to do with the Chern-Simons level. 
Some points on the $z$-plane are mapped as follows, 
\begin{equation}
z\ =\ 0,\ a,\ b,\ \infty \hspace{5mm} \mapsto \hspace{5mm} u\ = \ 0,\ K,\ K+iK',\ iK'
\end{equation}
where $K$ and $K'$ are the complete elliptic integrals 
\begin{equation}
K\ =\ K(k)\ =\ \int_0^1\frac{dx}{\sqrt{(1-x^2)(1-k^2x^2)}}, \hspace{5mm} K'\ =\ K\left( \sqrt{1-k^2} \right). 
\end{equation}

\begin{figure}[tbp]
\includegraphics{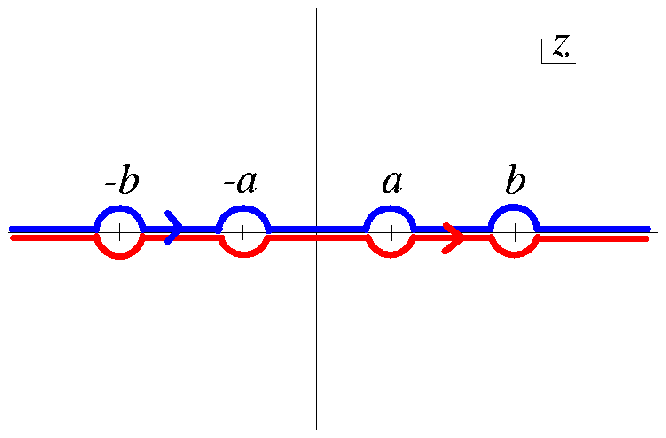}
\includegraphics{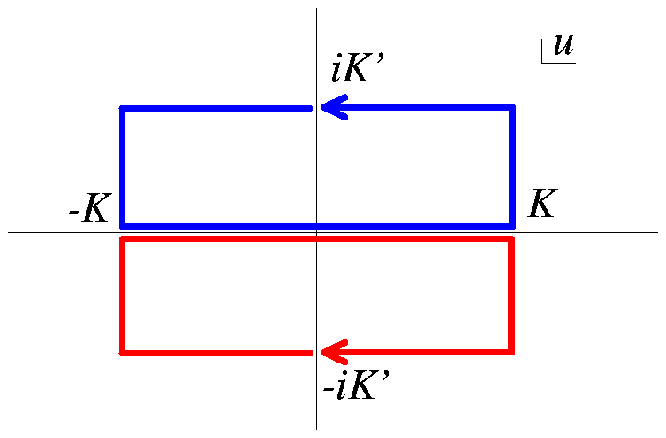}
\caption{
The conformal mapping from $z$-plane to $u$-plane. 
The upper-half $z$-plane is mapped into the region inside the blue rectangle in the $u$-plane, and the lower-half $z$-plane is mapped into 
the region inside the red rectangle in the $u$-plane.  
}
\label{map}
\end{figure}

The advantage of introducing $u$ is as follows. 
The inverse function of (\ref{u-def}) is 
\begin{equation}
u\ =\ \int_0^{z/a}\frac{dx}{\sqrt{(1-x^2)(1-k^2x^2)}}. 
\end{equation}
This defines a particular kind of the Schwarz-Christoffel mapping which maps the whole $z$-plane into a rectangular region $R_0$ 
where $R_l$ for an integer $l$ is defined as 
\begin{equation}
R_l\ :=\ \{\,u\in\mathbb{C}\ |\ (2l-1)K\le\mbox{Re}(u)\le (2l+1)K,\ -K'\le\mbox{Im}(u)\le K'\,\}. 
\end{equation}
Note that the upper and the lower sides of $[a,b]$ and $[-b,-a]$ are mapped on different regions. 
The mapping is depicted in Figure \ref{map}. 
An extension of $G(u):=G(z(u))$ from $R_0$ to the whole $u$-plane in an appropriate way determines $G(u)$ on the whole Riemann sheet. 

\vspace{5mm}

The function $G(u)$ satisfies some equations. 
The first one is the homogeneous equation (\ref{homo-appendix}). 
In terms of $u$, this is written as 
\begin{equation}
G(u+K)+G(-u+K)-2n\,G(u-K)\ =\ 0, 
   \label{1st}
\end{equation}
where $u\in i[0,K']$. 
The second one is the equation 
\begin{equation}
G(u-K)\ =\ G(-u-K), 
   \label{2nd}
\end{equation}
where $u\in i[0,K']$, requiring that $G(z)$ has no branch cut on $[-b,-a]$. 
The third one comes from the assumption that $G(z)$ is regular if $|\mbox{Re}(z)|>b$. 
This becomes 
\begin{equation}
G(u+iK')\ =\ G(u-iK'), 
   \label{3rd}
\end{equation}
where $u\in[-K,K]$. 

In \cite{Eynard:1995zv}, $G(u)$ is extended to the whole $u$-plane such that the above three equations are satisfied for {\it any} $u\in\mathbb{C}$. 
It is necessary to check whether such an extension can be done consistently. 

The first step is to extend $G(u)$ to the next region $R_{-1}$ using (\ref{2nd}). 
That is, the left-hand side of (\ref{2nd}) 
with $u\in K+R_0$, which is assumed to be already known, define $G(u)$ on $R_{-1}$ through the right-hand side. 

The next step is to use 
\begin{equation}
G(u-3K)-2n\,G(u-K)+G(u+K)\ =\ 0
\end{equation}
which is a consequence of (\ref{1st}) and (\ref{2nd}). 
This equation determines $G(u)$ on $R_l$ provided that $G(u)$ has been already defined on $R_{l-1}$ and $R_{l-2}$, or on $R_{l+1}$ and $R_{l+2}$. 
It is easy to check that this extension is compatible with (\ref{2nd}). 

Up to here, $G(u)$ has been defined on $\bigcup_{l\in\mathbb{Z}}R_l$. 
The extension to the vertical direction is straightforward by using (\ref{3rd}). 
Obviously, this last step is compatible with (\ref{1st}) and (\ref{2nd}). 

\vspace{5mm}

Now $G(u)$ is defined on the whole $u$-plane, satisfying (\ref{1st})(\ref{2nd})(\ref{3rd}) as functional equations, not boundary conditions. 
The next task is to solve these functional equations. 
It is convenient to define 
\begin{equation}
G_+(u)\ =\ \frac{e^{\frac{1}2\pi i\nu}G(u)+e^{-\frac{1}2\pi i\nu}G(-u)}{2\sin(\pi\nu)}. 
   \label{G_+ def}
\end{equation}
It can be shown that $G_+(u)$ satisfies the following equations 
\begin{eqnarray}
G_+(u+2K) &=& e^{-\pi i(1-\nu)}G_+(u), 
   \label{4th} \\
G_+(u+2iK') &=& G_+(u). 
   \label{5th}
\end{eqnarray}
The original function $G(u)$ can be recovered from $G_+(u)$ by 
\begin{equation}
G(u)\ = \ -i\left[ e^{\frac{1}2\pi i\nu}G_+(u)-e^{-\frac{1}2\pi i\nu}G_+(-u) \right]. 
\end{equation}
One can show that (\ref{1st})(\ref{2nd}) are consequences of (\ref{4th}) while (\ref{3rd}) is a consequence of (\ref{5th}). 
Therefore, the task is reduced to solving (\ref{4th})(\ref{5th}). 

\vspace{5mm}

There exists a solution 
\begin{eqnarray}
H_+(u)
&=& \frac{\vartheta_1(\frac{u-iK'}{2K})\vartheta_1(\frac{u-\varepsilon}{2K})}{\vartheta_1(\frac{u-K}{2K})\vartheta_1(\frac{u-(K+iK')}{2K})}
       e^{-\pi i(1-\nu)\frac u{2K}} \nonumber \\
&=& -i\frac{\vartheta_0(\frac{u}{2K})\vartheta_1(\frac{u-\varepsilon}{2K})}{\vartheta_2(\frac{u}{2K})\vartheta_3(\frac{u}{2K})}
       e^{-\pi i(1-\nu)\frac u{2K}}
\end{eqnarray}
of (\ref{4th})(\ref{5th}) 
where $\vartheta_a(v)=\vartheta_a\left( v,\frac{iK'}{K} \right)$ $(a=0,1,2,3)$ are the elliptic theta functions. 
Note that this solution differs from the one given in \cite{Eynard:1995zv} by the overall sign. 
One can easily verify that this is a solution of (\ref{4th})(\ref{5th}) provided that $\varepsilon$ is chosen to be 
\begin{equation}
\varepsilon\ :=\ i(1-\nu)K'. 
\end{equation}
Note that $H_+(u)$ has simple poles at $u=K$ and $u=K+iK'$ which correspond to $z=a$ and $z=b$, respectively. 
The conformal map (\ref{u-def}) maps near those points as follows, 
\begin{eqnarray}
u-K &\propto& {\sqrt{z-a}}, \\
u-(K+iK') &\propto& {\sqrt{z-b}}. 
\end{eqnarray}
Therefore, $H_+(z)$ has a branch cut on $[a,b]$ and it is divergent at the branch points. 
In addition, $H_+(z)$ has simple zeroes at $z=\infty$ ($u=iK'$) and $z=e$ ($u=\varepsilon$) where 
\begin{equation}
e\ :=\ a\,\mbox{sn}(i(1-\nu)K',k).
\end{equation}

In \cite{Eynard:1995zv}, $G_+(z)$ is chosen to be proportional to $H_+(z)$. 
The normalization condition chosen in \cite{Eynard:1995zv} is 
\begin{equation}
\lim_{z\to\infty}zG(z)\ =\ i. 
\end{equation}
This determines the normalization constant to be 
\begin{equation}
G_+(z)\ =\ a\frac{(\vartheta_3^0)^2}{\vartheta_0^0\vartheta_0(\frac\varepsilon{2K})}H_+(z). 
\end{equation}
This normalization implies the following product formula 
\begin{equation}
G_+(z)G_+(-z)\ =\ \frac{z^2-e^2}{(z^2-a^2)(z^2-b^2)}. 
\end{equation}

Note that $G_+(z)$ has also another branch cut on $[-b,-a]$ but it disappears in $G(z)$. 

\vspace{5mm}

The analysis in this paper requires to know the value $G_+(0)$. 
Using the identities of elliptic functions, one finds 
\begin{equation}
G_+(0)\ =\ ie.  
\end{equation}

\vspace{1cm}

\section{An identity} \label{identity}

\vspace{5mm}

In this Appendix, $ab=1$ is always assumed. 

Consider the following function, 
\begin{equation}
F(z)\ :=\ \tilde{g}(z^{-1})G_+(z^{-1}), \hspace{5mm} \tilde{g}(z)\ :=\ zg(z), 
\end{equation}
where $g(z)$ is defined in (\ref{g(z)-def}). 
$F(z)$ has singularities at $z=a,b$ just as $G_+(z)$. 
One can show that $\tilde{g}(z^{-1})$ has a zero at $z=e$ while $G_+(z^{-1})$ is finite there, implying that $F(z)$ has a zero at $z=e$. 
$F(z)$ behaves as 
\begin{equation}
F(z)\ :=\ \frac i{ez}+O(z^{-2})
   \label{F-infty}
\end{equation}
near infinity. 

The product $F(z)F(-z)$ turns out to have a simple form: 
\begin{eqnarray}
F(z)F(-z) 
&=& -z^{-2}\frac{(z^{-2}-a^2)(z^{-2}-b^2)-\frac{\bar{e}^2}{e^2}z^{-2}}{(z^{-2}-e^2)^2}G_+(z^{-1})G_+(-z^{-1}) \nonumber \\
&=& -z^{-2}\frac{e^2(z^{-2}-a^2)(z^{-2}-b^2)-(e^2-a^2)(e^2-b^2)z^{-2}}{e^2(z^{-2}-e^2)(z^{-2}-a^2)(z^{-2}-b^2)} \nonumber \\
%&=& -z^{-2}\frac{e^2z^{-2}-1}{e^2(z^{-2}-a^2)(z^{-2}-b^2)} \nonumber \\
&=& \frac{z^2-e^2}{e^2(z^2-a^2)(z^2-b^2)}. 
\end{eqnarray}
This implies that $z=a,b$ are the only singularities of $F(z)$, and $z=e,\infty$ are the only zeroes of $F(z)$. 

In term of the $u$ variable defined in (\ref{u-def}), $F(z)$ is written as 
\begin{equation}
F(u)\ =\ \tilde{g}(u+iK')G_+(u+iK'). 
\end{equation}
It is shown in Appendix \ref{G(z)} that $G_+(u)$ is defined on the whole $u$-plane and it satisfies the relations (\ref{4th})(\ref{5th}). 
Therefore, if $\tilde{g}(u)$ satisfies 
\begin{equation}
\tilde{g}(u+2K)\ =\ \tilde{g}(u), \hspace{5mm} \tilde{g}(u+2iK')\ =\ \tilde{g}(u) 
   \label{tilde{g}}
\end{equation}
for any $u\in\mathbb{C}$, then, since the ratio $F(z)/G_+(z)$ turns out to be an elliptic function without poles, 
it is concluded that $F(z)$ is proportional to $G_+(z)$. 

One can show that 
\begin{equation}
\tilde{g}(-y\pm i0)\ =\ \tilde{g}(y\pm i0). 
\end{equation}
In terms of the $u$ variable, this can be written as 
\begin{equation}
\tilde{g}(u-K)\ =\ \tilde{g}(u+K). \hspace{5mm} (u\in i[-K',K])
\end{equation}
In addition to this, one has the following condition 
\begin{equation}
\tilde{g}(u+iK')\ =\ \tilde{g}(u-iK'), \hspace{5mm} (u\in[-K,K])
\end{equation}
which comes from the regularity outside the branch cuts. 
These conditions imply that $\tilde{g}(u)$ can be easily extended to the whole $u$-plane while satisfying (\ref{tilde{g}}) 
for any $u\in\mathbb{C}$. 

The proportionality constant is determined from the behavior (\ref{F-infty}). 
This completes the proof of the identity 
\begin{equation}
e\,g(z^{-1})G_+(z^{-1})\ =\ zG_+(z). 
\end{equation}

\vspace{1cm}

\section{Uniqueness of the solution} \label{uniqueness}

\vspace{5mm}

This appendix discusses conditions which uniquely determine the solution of the equation 
\begin{equation}
f(y) \ =\ v(y+i0)+v(y-i0)-2n\,v(-y), 
   \label{SP-app}
\end{equation}
where $y\in[a,b]\subset\mathbb{R}$. 
The function $f(y)$ can be arbitrary. 

\vspace{5mm}

Suppose that there are two solutions $v_1(z)$ and $v_2(z)$ of (\ref{SP-app}). 
The difference $\delta v(z):=v_1(z)-v_2(z)$ satisfies the homogeneous equation 
\begin{equation}
0\ =\ \delta v(y+i0)+\delta v(y-i0)-2n\,\delta v(-y). 
   \label{homo-app}
\end{equation}
The general solution of this equation is given in subsection \ref{subsection homo}. 
The result is 
\begin{equation}
\delta v_+(z)\ =\ \Bigl[ A(z^2)+zB(z^2)\sqrt{(z^2-a^2)(z^2-b^2)} \Bigr]G_+(z), 
\end{equation}
where $\delta v_+(z)$ is defined as in (\ref{omega_+-def}). 

Now, we require that 
\begin{itemize}
\item $v_1(z), v_2(z)$ are holomorphic on $\mathbb{C}\backslash[a,b]$, and 
\item $v_1(z), v_2(z)$ are finite at $z=a,b$. 
\end{itemize}
Then, as shown in subsection \ref{subsection homo}, $\delta v_+(z)$ should be of the form 
\begin{equation}
\delta v_+(z)\ =\ \Bigl[ \tilde{A}(z^2)g(z)+z\tilde{B}(z^2) \Bigr]\sqrt{(z^2-a^2)(z^2-b^2)}G_+(z), 
\end{equation}
where $g(z)$ is defined as (\ref{g(z)-def}). 
Both $\tilde{A}(z^2)$ and $\tilde{B}(z^2)$ are entire functions on $\mathbb{C}$. 
A further requirement 
\begin{itemize}
\item $v_1(z), v_2(z)$ are finite at infinity
\end{itemize}
demands that $\tilde{A}(z^2)$ and $\tilde{B}(z^2)$ must vanish at infinity. 
Therefore, one finds 
\begin{equation}
\tilde{A}(z^2)\ =\ \tilde{B}(z^2)\ =\ 0. 
\end{equation}
Since $\delta v_+(z)=0$ implies $\delta v(z)=0$, it is concluded that the above three conditions determine the solution of (\ref{SP-app}) 
uniquely.

\vspace{1cm}

\section{Determination of coefficients} \label{coeff}

\vspace{5mm}

The residues of $\omega_+(z,\xi)$ at $z=\pm\xi$ are 
\begin{eqnarray}
\mbox{Res}_\xi\,\omega_{+}(z,\xi) &=& \bigl( c(\xi)g(\xi)+d(\xi)\xi \bigr)\tilde{G}_+(\xi), \\
\mbox{Res}_{-\xi}\,\omega_+(z,\xi) &=& \bigl( c(\xi)g(-\xi)-d(\xi)\xi \bigr)\tilde{G}_+(-\xi), 
\end{eqnarray}
where $\tilde{G}_+(z)$ is defined as 
\begin{equation}
\tilde{G}_+(z)\ :=\ \frac{\sqrt{(z^2-a^2)(z^2-b^2)}}{2z}G_+(z). 
\end{equation}
The coefficients $c(\xi)$ and $d(\xi)$ are the solution of 
\begin{equation}
\left[ 
\begin{array}{cc}
g(\xi)\tilde{G}_+(\xi) & \xi\tilde{G}_+(xi) \\
g(-\xi)\tilde{G}_+(-\xi) & -\xi\tilde{G}_+(-\xi) 
\end{array}
\right]\left[ 
\begin{array}{c}
c(\xi) \\ d(\xi)
\end{array}
\right]\ = \ -\frac {i}{2(n^2-1)}\left[ 
\begin{array}{c}
e^{-\frac12{\pi i\nu}} \\ 
e^{\frac12{\pi i\nu}}
\end{array}
\right]. 
\end{equation}
Let the above $2\times2$ matrix be denoted by $A$. 
Its determinant is 
\begin{eqnarray}
\det A
&=& -\xi\bigl( g(\xi)+g(-\xi) \bigr)\tilde{G}_+(\xi)\tilde{G}_+(-\xi) \nonumber \\
&=& \frac{\sqrt{(\xi^2-a^2)(\xi^2-b^2)}}{2\xi}. 
\end{eqnarray}
and therefore, the inverse is 
\begin{eqnarray}
A^{-1}
&=& \frac{2\xi}{\sqrt{(\xi^2-a^2)(\xi^2-b^2)}}\left[ 
\begin{array}{cc}
-\xi\tilde{G}_+(-\xi) & -\xi\tilde{G}_+(\xi) \\
-g(-\xi)\tilde{G}_+(-\xi) & g(\xi)\tilde{G}_+(\xi) 
\end{array}
\right] \nonumber \\
&=& \left[ 
\begin{array}{cc}
\xi G_+(-\xi) & -\xi G_+(\xi) \\
g(-\xi)G_+(-\xi) & g(\xi)G_+(\xi) 
\end{array}
\right]. 
\end{eqnarray}
The solutions is then  
\begin{eqnarray}
c(\xi)  
&=& -\frac i{2(n^2-1)}\xi\left( e^{-\frac{\pi i\nu}2}G_+(-\xi)-e^{\frac{\pi i\nu}2}G_+(\xi) \right) 
      \nonumber \\
&=& -\frac{1}{2(n^2-1)}\xi G(\xi). \\
d(\xi) 
&=& -\frac{i}{2(n^2-1)}\left( e^{-\frac{\pi i\nu}2}g(-\xi){G}_+(-\xi)+e^{\frac{\pi i\nu}2}g(\xi){G}_+(\xi) \right) \nonumber \\
&=& -\frac{i}{2(n^2-1)}\frac{1}{e}\left( -e^{-\frac{\pi i\nu}2}\xi^{-1}{G}_+(-\xi^{-1})+e^{\frac{\pi i\nu}2}\xi^{-1}{G}_+(\xi^{-1}) \right) \nonumber \\
&=& \frac{1}{2(n^2-1)e}\xi^{-1}G(\xi^{-1}). 
\end{eqnarray}

\end{document}